\documentclass[11pt]{article}

\usepackage{setspace}

\usepackage{latexsym}
\usepackage{algorithm}
\usepackage[xdvi,dvips]{graphics}
\usepackage{pstricks}
\usepackage{epsfig}
\usepackage{amssymb}
\usepackage{multicol}
\usepackage{array}
\usepackage{float}
\usepackage{boxedminipage}

\newenvironment{proof}{{\bf Proof. } }{{\hfill $\Box$}\vspace{.5pc}}
\newtheorem{theorem}{Theorem}[section]
\newtheorem{corollary}[theorem]{Corollary}
\newtheorem{definition}[theorem]{Definition}
\newtheorem{lemma}[theorem]{Lemma}

\newtheorem{proposition}[theorem]{Proposition}

\floatstyle{ruled}

\newfloat{algo}{thp}{lop}[section]
\floatname{algo}{Algorithm}

\newcommand{\BEGLIST}{\begin{list}{}{\partopsep -2pt \parsep -2pt \listparindent 0pt}}% \labelwidth .5in}}
\newcommand{\ENDLIST}{\end{list}}

\newcommand{\mN}   {\mathcal{N}}

\newcommand{\mP}   {\mathcal{P}}

\newcommand{\mX}   {\mathcal{X}}
\newcommand{\bbZ}   {\mathbb{Z}}

\newcommand{\TRUE}{\mathtt{true}}
\newcommand{\FALSE}{\mathtt{false}}

\newcommand{\eg}{\emph{e.g., }}
\newcommand{\ie}{\emph{i.e., }}

\newcommand{\text}[1]{\mbox{#1}}

\begin{document}

%\title{ Self-Stabilizing Wavelet Stream and $\varrho$-Distance Synchronization}
\title{ Self-Stabilizing Wavelets and $\varrho$-Hops Coordination}
%\small{Extended Abstract}

\author{
Christian Boulinier and Franck Petit\\
LaRIA, CNRS \\
Universit\'{e} de Picardie Jules Verne, France\\
}

\date{}
\maketitle

\footnotesize
\begin{abstract}
We introduce a simple tool called the \emph{wavelet} (or, $\varrho$-wavelet) scheme.  
Wavelets deals with coordination among processes which are at most
$\varrho$ hops away of each other.  
We present a self-stabilizing solution for this scheme.  Our solution requires no underlying 
structure and works in arbritrary anonymous networks, \ie no process identifier is required.  
Moreover, our solution works under any (even unfair) daemon. 

Next, we use the wavelet scheme to design self-stabilizing \emph{layer clocks}.  
We show that they provide an efficient device in the design of local coordination 
problems at distance $\varrho$, \ie $\varrho$-barrier synchronization and
$\varrho$-local resource allocation (LRA) such as $\varrho$-local mutual exclusion (LME), 
$\varrho$-group mutual exclusion (GME), and $\varrho$-Reader/Writers.  Some solutions to 
the $\varrho$-LRA problem (\eg $\varrho$-LME) also provide transformers 
to transform algorithms written assuming any $\varrho$-central daemon into algorithms working 
with any distributed daemon.  
\\

\end{abstract}
\normalsize

%\thispagestyle{empty}

%\baselineskip 5.0mm \-\unitlength=1mm
%
%\bigskip {\bf Corresponding Author}:
%
%\begin{quotation}
%Franck Petit
%
%Email: Franck.Petit@u-picardie.fr
%
%LaRIA, Universit\'{e} de Picardie Jules Verne, Amiens, France
%
%Tel. +33-322-828-878
%
%\end{quotation}
%
%\newpage 

%\setcounter{page}{1}
%\doublespacing
%\onehalfspacing
%\singlespacing
\section{Introduction}

%\cite{ABDT98}  \cite{Gar03} \cite{HL01}
Most of the distributed system are not fully connected networks.  
Each process is only directly connected with a subset of others, called neighbors.  By this way, 
the communication links organize the network in a some graph topology which is either arbitrary or 
in accordance with some global topology constraints, \eg acyclicity, constant degree, ring, grid, etc.  
Whatever the topology complexity, the design 
of a distributed task is simplified if it requires coordination mechanisms involving a process with
its neighbors only, \ie  %is required to synchronize have depends on 
%its neighbors only, \ie 
one hop away.  Such distributed tasks are said to be \emph{local}.  
Unfortunately, many distributed tasks requires coordination farther away than the immediate neighbors, 
\ie $\varrho$ hops away with $\varrho > 1$.  If $\varrho$ is equal to the diameter of the network $D$, then
the task is said to be \emph{global}. 

In this paper, we consider problems requiring coordination among processes which are 
at most $\varrho$ hops away of each other. We present solutions having the desirable property of 
self-stabilization.  The concept of {\em self-stabilization}~\cite{D74,D00} is an efficient 
approach to design distributed systems to tolerate arbitrary transient faults. A
self-stabilizing system, regardless of the initial states of the processors
and initial messages in the links, is guaranteed to converge to the intended
behavior in finite time.

\paragraph{Motivation and Related Works.}

Coordination at distance $\varrho$ received a particular attention in recent works.  
There are various motivations for this issue.  The first one consists in the design of $\varrho$-local
computations~\cite{NS93}, \ie running in constant time independent of any global parameter like 
the size of the network or the diameter.  Computation in constant time $\varrho$ can be achieved if the processes
can collect informations from processes located within radius of $\varrho$ from them. 
In \cite{NS93}, the authors mainly address \emph{Local Checkable Labeling} problems. 
Local computation is also considered in~\cite{GMM04} by considering the recognition problem.  
The computing model is a relabelling system.  
%In this paper, local means that each relabelling step changes only the labels
%of a connected subgraph of a fixed radius  $\varrho$.  The change of the labels
%depends on the local context of the subgraph. The relabelling relation allows 
%parallel relabelling since non-overlapping balls may be relabelled independently. 

Wireless networks bring new trends in distributed systems which also motivate research the local control 
of concurrency at distance $\varrho$.  In~\cite{DNT06}, the authors propose a generalization of the 
well-known dining philosophers problem~\cite{Dij68}.  They extends the conflict processes beyond the 
immediate neighbors of the processes.  As an application, their solution provide a solution
to the interfering transmitter problem in wireless networks.

Another motivation consists in assuming that the knowledge of the processes goes beyond their immediate neighbors 
could help in the design of non-trivial tasks~\cite{GGHK04,GHJT06}. 
An efficient self-stabilizing solution is given to the \emph{maximal 2-packing} problem assuming
the knowledge at distance $2$ \cite{GGHK04}. (The maximal 2-packing problem consists to find a 
maximal set of nodes $S$, such that no two nodes in $S$ are adjacent and no two nodes in $S$ have a common neighbor.)
The solution in \cite{GGHK04} requires process ID's and works under a central daemon.  
In \cite{GHJT06}, the authors propose a $\varrho$-distance knowledge transformer to
construct self-stabilizing algorithms which use a $\varrho$ distance 
knowledge. Again, their solution works only if the daemon is central and with process ID's.

Note that various kinds of transformers have been proposed in the area of 
self-stabilization to refine self-stabilizing algorithms which use tight
scheduling constraints like the central daemon into the corresponding
self-stabilizing algorithm working assuming weaker daemons,\eg 
\cite{MN98,GH99,NA02,CDP03}. A popular technique consists in composing the 
algorithm with a self-stabilizing \emph{local mutual exclusion} (LME) 
algorithm~\cite{MN98,GH99,NA02}. 
LME allows to grant privileged processes to enter critical section if and only 
if none of their neighbors has the privilege, this infinitely often. 
So, Any LME-based solution does not allow concurrent execution of neighboring processes. 
The solution in~\cite{CDP03} is based on the \emph{Local Resource Allocation} (LRA), which 
allows neighboring processes to enter their critical sections concurrently provided
they do not use conflicting resources.
It transforms any algorithms written in a high-atomicity model (\eg with a central daemon)
into the distributed read/write atomicity model by allowing neighborhood concurrency.

However, none of the above solutions allows coordination farther than the immediate neighbors.
So, they are not directly applicable to the method developed in \cite{NS93,GGHK04,GHJT06,DNT06}.

\paragraph{Contributions.}

In this paper, we introduce a simple tool called the \emph{wavelet} (or, $\varrho$\emph{-wavelet}) scheme.  
Wavelets deals with coordination among processes which are at most $\varrho$ hops away of each other.  
Wavelets are related to the notion of \emph{wave} (also called \emph{Total} algorithm~\cite{T88,Tel94}).

In this paper, we present a self-stabilizing solution to the $\varrho$-wavelet problem.
There are several way to design the wavelet scheme depending on network properties.   
For instance, assuming a unique identifier on each process, in~\cite{DNT06}, the authors 
provides a self-stabilizing $\varrho$-wavelets scheme by combining a stabilizing Propagation of 
Information with Feedback (PIF)~\cite{BDPV99b} over a self-stabilizing BFS spanning tree~\cite{HC92,J97} 
rooted at each process of height equal to $\varrho$.  

By contrast, our solution requires no underlying structure and works in arbritrary anonymous networks, \ie
no process identifier is required.
Our solutions is based on the unison in~\cite{BPV04b} and works assuming any distributed (even unfair) daemon. 

Next, we use the wavelet scheme to design self-stabilizing \emph{layer clocks}.
The lower layer clock, in the sequel called the \emph{main} clock, provides a wavelet stream. 
The upper layer clock, so called the \emph{slave} clock, achieves a $\varrho-$\emph{barrier} synchronization
mechanism, where no process $p$ starts 
to execute its phase $i+1$ before all processes in the $\varrho-$ball centered in $p$ have completed their phase $i$.

Finally, we show that the layer clock also provides an efficient underlying device in the design of 
various local resource allocation problems at distance $\varrho$.  This problems include
Mutual Exclusion~\cite{D65}, Group Mutual Exclusion~\cite{Jou00}, and Readers-Writers~\cite{CHP71}.
Some of these solutions (\eg $\varrho$-LME) also provides transformers 
to transform algorithms written assuming any $\varrho$-central daemon into algorithms working 
with any distributed daemon.

\paragraph{Paper Outline.}

The remainder of the paper is organized as follows. 
We formally describe notations, definitions, and the execution model in Section~\ref{sec:prel}. We also 
state what it means for a protocol to be self-stabilizing. 
In Section~\ref{sec:wave}, we define the wavelet scheme, present our solution for this problem in an arbitrary anonymous networks, 
and show how it can be used as an infimum computation at distance $\varrho$. 
In Section~\ref{sec:distance_scheme},  we introduce the self-stabilizing layer clocks and show how they can be used 
to solve $\varrho$-local coordination problems. Finally, we make some concluding remarks in Section~\ref{sec:conclusion}.

\section{Preliminaries}
\label{sec:prel}

In this section, we first define the model of distributed systems 
considered in this paper.  We then define the execution model and various general definitions such as events, causal DAG, and Coherent Cuts.
We also state what it means for a protocol to be self-stabilizing.  

\subsection{Distributed System}

A \emph{distributed system} is an undirected connected graph, $G=(V,E)$,
where $V$ is a set of nodes---$|V|=n,\ n \geq 2$---and $E$ is the set of edges. Nodes
represent \emph{processes}, and edges represent \emph{bidirectional
communication links}. A communication link $(p,q)$ exists iff $p$ and
$q$ are neighbors.
The distributed system is considered to be arbitrary and anonymous, \ie we consider no particular topology 
nor unique identifiers on processes.

The set of neighbors of every process $p$ is denoted as $\mN_p$.
The \emph{degree} of $p$ is the number of neighbors of $p$, \ie equal to $|\mN_p|$.
The distance between two processes $p$ and $q$, denoted by $d\left( p,q\right) $,
is the length of the shortest path between $p$ and $q$.  Let $\varrho$ be a positive integer. Define $V(p,\varrho)$ 
as the set of processes such that $d(p,q) \leq \varrho$, \ie the $\varrho-$ball centered at $p$. 
%following subset of processes: be the If $k$ is an integer, we define 
%$D\left( p,k\right)=_{def}\left\{ q\in V,d\left( p,q\right) \leq k\right\} $. 
 $D$ denote the diameter of the network.

The program of a process consists of a set %$\mR_p$ 
of registers (also referred to as variables) 
and a finite set %$\mA_p$ 
of guarded actions of the following form: 
$<label>::\ <guard>\ \longrightarrow <statement>$.
Each process can only write to 
its own registers, and read its own registers and registers owned by the neighboring processes.
The guard of an action in the program of $p$ is a boolean
expression involving the registers of $p$ and its neighbors. The
statement of an action of $p$ updates one or more registers of $p$. 
An action can be executed only if its guard evaluates to true.   
The actions are atomically executed, meaning the evaluation of a guard and the execution of
the corresponding statement of an action, if executed, are done in one atomic step.

\subsection{Execution Model}

The \emph{state} of a process is defined by the values of its registers.
The \emph{configuration} of a system is the product of the states of all processes. 
Let a distributed protocol $\mP$ be a collection of binary transition
relations denoted by $\mapsto $, on $\Gamma$, the set of all
possible configurations of the system.  $\mP$ describes an
oriented graph $S=(\Gamma, \mapsto)$, called the \emph{transition graph} of $\mP$.
A sequence $e=\gamma_0, \gamma_1, \ldots, \gamma_i,\gamma_{i+1},\ldots$ is called an
\emph{execution} of $\mP$ iff $\forall i\geq 0, \gamma_{i}\mapsto \gamma _{i+1} \in S$. 
A process $p$ is said to be \emph{enabled} in a configuration
$\gamma_i \;( \gamma _i\in \Gamma)$ if there exists an action $A$ such that 
the guard of $A$ is true in $\gamma_i$.  
%When there is no ambiguity, we will omit $i$.
Similarly, an action $A$ is said to be enabled (in $\gamma$) at $p$
if the guard of $A$ is true at $p$ (in $\gamma$).
We consider that any enabled processor $p$ is \emph{neutralized} 
in the computation step $\gamma_i \mapsto \gamma_{i+1}$ if $p$ is enabled in 
$\gamma_i$ and not enabled in $\gamma_{i+1}$, but does not execute any action
between these two configurations.  (The neutralization of a processor represents the
following situation: At least one neighbor of $p$ changes its state between 
$\gamma_i$ and $\gamma_{i+1}$, and this change effectively made the guard of 
all actions of $p$ false.)

We assume an \emph{unfair and asynchronous distributed daemon}. 
\emph{Unfairness} means that even if a 
processor $p$ is continuously enabled, then $p$ may never be chosen by 
the daemon unless $p$ is the only enabled processor.
The \emph{asynchronous distributed} daemon implies that 
during a computation step, if one or more processors are enabled, then the 
daemon chooses at least one (possibly more) of these enabled processors to 
execute an action.

%If is not required to be \emph{fair}, \ie even if a 
%process $p$ is continuously enabled, then $p$ may never be chosen by 
%the daemon unless $p$ is the only enabled process.

In order to compute the time complexity, we use the definition of
\emph{round}~\cite{DIM97a}.  This definition captures the execution rate of 
the slowest processor in any computation.
Given an execution $e$, %($e \in \mathcal{E}$)
 the \emph{first round} of $e$ 
(let us call it $e^{\prime}$)
is the minimal prefix of $e$ containing the execution of one action 
of the protocol or the neutralization of every enabled processor from the first configuration.  
Let $e^{\prime \prime}$ be the suffix of $e$, \ie $e=e^{\prime}e^{\prime \prime}$.  
Then \emph{second round} of $e$ is the first round of $e^{\prime \prime}$, and so on.

\subsection{Events, Causal DAG's and Cuts}

\begin{definition}[Events] 
Let $\gamma _0\gamma _1....$ be a finite or
infinite execution. 
 For all $ p\in V,\left( p,0\right) $ is an event.
  Let $\gamma _{t}\rightarrow \gamma _{t+1}$ be a transition. If the
process $p$ executes a guarded action during this transition, we say that $p$
executes an action at time $t+1$.  The pair $\left( p,t+1\right)$ is said to be an
event (or a $p$-event). 
Events so that the guard does not depend on the shared registers of any neighbor are said to be \emph{internal}.
\end{definition}

\begin{definition}[Causal DAG]
The causal DAG associated is the smallest relation $\leadsto $ on the set of events
such that the following two conditions hold:
\BEGLIST
%\begin{enumerate}
\item [1.] Let $( p,t)$ and $(p,t')$ be two events such that $t>t_0$, $t'$ is 
the greatest integer such that $t_0 \leq t' < t$. Then,
$\left( p,t^{\prime }\right) \leadsto \left( p,t\right)$;

\item [2.] 
Let $(p,t)$ and $(q,t')$ be two events such that $(p,t)$ is not an internal event, $q \in \mN_p$, $t>t_0$, and $t'$ is 
the greatest integer such that $t_0 \leq t' < t$.  Then, 
$\left( q,t^{\prime }\right) \leadsto \left( p,t\right) $.
\ENDLIST
%\end{enumerate}
\end{definition}

Denote the \emph{causal order} on the sequence $\gamma _0\gamma _1....$ by $\preceq$.  Relation $\preceq$ is 
the reflexive and transitive closure of the causal
relation $\leadsto $.
The \emph{past cone} of an event $\left( p,t\right) $ is the  causal-$DAG$ induced
by every event $\left( q,t^{\prime }\right) $ such that $\left( q,t^{\prime
}\right) \preceq \left( p,t\right) $. 
A past cone involves a process $q$ iff there is a $q$-event in the cone.  We say that a past 
cone \emph{covers} $V$, iff every process $q\in V$ is involved in the cone. The \emph{cover} of an 
event $(p,t)$, denoted by $Cover(p,t)$, is the set of processes $q$ covered by the past cone of $(p,t)$. 

%If $\left( q,t^{\prime }\right) \preceq \left( p,t\right) $ then there
%exists a \emph{causality chain } from $\left( q,t^{\prime }\right) $ to $(p,t)$:  $\left( q,t^{\prime }\right) =\left(
%q_{0},t_{0}\right) \leadsto $ $\left( q_{1},t_{1}\right) \leadsto $ $\left(
%q_{2},t_{2}\right) ...\leadsto $ $\left( q_{r},t_{r}\right) =\left(
%p,t\right) $ , its   associated \emph{walk} is the \emph{walk} $q_{0}q_{1}$...$q_{r}$. The \emph{walk cover} of an event $(p,t) $ is the set of \emph{walks} associated to  the causality chains ending to $(p,t)$.    This set is denoted by $WalkCover(p,t)$.  Of course, this set contains the \emph{walk} of length $0$ denoted by $p$. 

\begin{definition}[Cut]
A cut $C$ on a causal DAG is a map from $V$ to $\Bbb{N}$, which associates
a process $p$ with a time $t_{p}^{C}$. We mix this map with its graph: $C=
\left\{ \left( p,t_{p}^{C}\right) ,p\in V\right\} $.%\newline
\end{definition}

The \emph{past} of $C$, denoted by $\left]\leftarrow ,C\right]$, is the set of events 
$(p,t)$ such that $t\leq t_{p}^{C}$. 
Similarly, we define the \emph{future} of $C$, denoted by $\left[ C,\rightarrow \right[$, as the set of 
events $(p,t)$ such that $t_{p}^{C}\leq t$. 
A cut is said to be \emph{coherent} if $\left( q,t^{\prime }\right) \preceq \left(
p,t\right) $ and $\left( p,t\right) \preceq \left( p,t_{p}^{C}\right)$,
then $\left( q,t^{\prime }\right) \preceq \left( q,t_{q}^{C}\right)$ .
%\newline
A cut $C_{1}$ is less than or equal to a cut $C_{2}$, denoted by $C_{1}\preceq C_{2}$,
if the past of $C_{1}$ is included in the past of $C_{2}.$%\newline

If $C_{1}$ and $C_{2}$ are coherent cutes such that $C_{1}\preceq C_{2}$, then $\left[ C_{1},C_{2}\right] $ is the 
\emph{induced} causal DAG defined by the events $\left( p,t\right) $ such
that $\left( p,t_{p}^{C_{1}}\right) \preceq \left( p,t\right) \preceq \left(
p,t_{p}^{C_{2}}\right) $.%\newline
A \emph{sequence of events} is any segment $\left[ C_{1},C_{2}\right] $
where $C_{1}$ and $C_{2}$ are coherent cuts satisfying $C_{1}\preceq C_{2}$.  
Any event of $C_1$ is called an \emph{initial event}.

\subsection{Self-Stabilization}
Let $\mX$ be a set. A \emph{predicate} $P$ is a function that has a Boolean 
value---$\TRUE$ or $\FALSE$---for each element $x\in \mX$.
A predicate $P$ is \emph{closed} for a transition graph $S$ iff 
every state of an execution $e$ that starts in a state satisfying $P$ also satisfies $P$.
A predicate $Q$ is an attractor of the predicate $P$, denoted by $P \vartriangleright Q$,
iff $Q$ is closed for $S$ and for every execution $e$ of $S$, beginning by a state satisfying $P$, 
there exists a configuration of $e$ for which $Q$ is true. 
A transition graph $S$ is \emph{self-stabilizing} for a predicate $P$ iff $P$ is an attractor 
of the predicate $\TRUE$, \ie $\TRUE \vartriangleright P$.

\section{Wavelets}
\label{sec:wave}

In this section, we first define the problem considered in this paper, followed 
by our self-stabilizing solution designed for any anonymous networks. Next, we show that 
it provides an efficient tool to compute any infimum at distance $\varrho$. 

\subsection{Problem Definition}
%\begin{definition}[ Wavelet ]
%\label{def:wave}
Let us assume that there exists a special internal type of events called a \emph{decide} event. Let  $\varrho$ be an integer.
A \emph{$\varrho$-wavelet} is a \emph{sequence of events}  $\left[ C_{1},C_{2}\right]$
that satisfies the following two requirements:

\BEGLIST
%\begin{enumerate}
\item [1.] The causal DAG induced by $\left[ C_{1},C_{2}\right] $ contains at least one decide event;

\item [2.] For each decide event $\left( p,t\right)$, the past of $\left( p,t\right) $ in 
$\left[ C_{1},C_{2}\right] $ covers $ V(p,k)$.
%\end{enumerate}
\ENDLIST
%\end{definition}

A \emph{wave} is the particular case where $\varrho \geq D$, $D$ is the diameter of the network.
There are several way to implement the $\varrho$-wavelet scheme if the processes have Id's, for instance using
the PIF scheme on trees with height equal to $\varrho$ rooted at each process.  In the following subsection,
we present a solution for the $\varrho$-wavelet problem in anonymous networks.  Next, we show how this solution 
provides a self-stabilizing infimum computation in a $\varrho$-ball.

%\begin{lemma}
%Let $\left[ C_{1},C_{2}\right] $ be a wave or a \emph{BLF}. For each $p \in V$, there exists a unique $p$-event in $\left[ C_{1},C_{2}\right] $
%which is minimal among the p-events in $\left[ C_{1},C_{2}\right] $, according to the order $\preceq $. We call it the initial event of $p$ in $\left[ C_{1},C_{2}\right] $.

%\end{lemma}
%\begin{proof}
%Let $(q,t)$ a decide event, its past in $\left[ C_{1},C_{2}\right] $ covers $V$, so there exists a $p$-event. 
%The set of $p$-events is a finite chain, according to $\preceq $. The lemma follows.
%\end{proof}

\subsection{Solution Description}

Our solution is based on the unison developed in~\cite{BPV04b}, which  stabilizes in 
$O(n)$ rounds in general graphs. 
Note that in a tree, we could use the protocol proposed in~\cite{BPV06}.  It gives the better 
stabilization time complexity of at most $D$ rounds.
In the sequel, we first borrow some basic definitions and properties introduced in~\cite{BPV04b},
followed by our solution and its correctness proof. 

\subsubsection{Unison}
\label{sub:unison}

%We first borrow some basic definitions and properties introduced in~\cite{BPV04b}. 

\paragraph{Basic Definitions and Properties.}

Let $\bbZ$ be the set of integers and $K$ be a strictly positive integer.
Two integers $a$ and $b$ are said to be \emph{congruent modulo} $K$, denoted by
$a\equiv b [K]$ if and only if $\exists \lambda \in \bbZ,\
b = a + \lambda K$. We denote $\bar{a}$ the unique element in $[0,K-1]$
such that $a \equiv \bar{a} [K]$.  $\min ( \overline{a-b}, \overline{b-a})$ is a \emph{distance} 
on the torus $[0,K-1]$ denoted by $d_K (a,b)$  . 
Two integers $a$ and $b$ are said to be \emph{locally comparable}  if and only if
 $d_K (a,b) \leq 1$.  We then define
the \emph{local order relationship} $\leq_l$ as follows: 
$
a\leq_l b \stackrel{\mathrm{def}}{\Leftrightarrow} 0 \leq \overline{b-a} \leq 1
$.
If $a$ and $b$ are two locally comparable integers,  we define  
$b \ominus a $ as follows:
$
b \ominus a \stackrel{\mathrm{def}}{=} \mbox{  if } a \leq _l b \mbox{ then } \overline{b-a} \mbox{ else }  - \overline{a-b}    
$.
If $a_0, a_1, a_2,\ldots a_{p-1}, a_p$ is a sequence of integers such that 
$\forall i \in \{0,\ldots,p-1\}$, $a_i$ is locally comparable to $a_{i+1}$,
then $S=\sum \limits_{i=0}^{p-1}\left( a_{i+1} \ominus a_i\right) $ is the \emph{local variation} of this sequence. 
%Clearly, $S \equiv a_p - a_0 [K]$.% and 
%$S \equiv 0 [K] \Leftrightarrow a_p - a_0 \equiv 0[K].$ 
 
%\paragraph{Incrementing System.} We D
Define $\mX = \{-\alpha,\ldots,0,\ldots,K-1\}$, where $\alpha$ is a positive integer. 
Let $\varphi$ be the function from $\mX$ to $\mX$ defined by:
%$$ 
%\varphi: x \rightarrow 
%\left\{ 
%  \begin{array}{ll}
%     \overline{x+1} & \mbox{if } x \geq 0 \\
%     x+1            & \mbox{otherwise}
%  \end{array}
%\right. 
%$$
$
\varphi (x) \stackrel{\mathrm{def}}{=} \mbox{  if } x \geq 0 \mbox{ then } \overline{x+1} \mbox{ else }  x+1    
$.
The pair $(\mX,\varphi)$ is called a \emph{finite incrementing system}. %---refer to Figure %~\ref{fig:cerise}.
%
%\begin{figure}[!htbp]
%\begin{center}
%    \epsfig{file=cerise.eps, width=0.35\linewidth}\\
%\end{center}
%\caption{The finite incrementing system $(\mX,\varphi)$.}
%\label{fig:cerise}
%\end{figure}
$K$ is called the \emph{period} of $(\mX, \varphi)$.
%Let $\alpha = -\kappa$ be the \emph{initial element} of $(\mX,\varphi)$.
Let $tail_{\varphi}=\{-\alpha,\ldots,0\}$ and $ring_{\varphi}=\{0,\ldots,K-1\}$ be the sets of 
``extra'' values and ``expected'' values, respectively.
The set $tail_\varphi ^*$ is equal to $tail_\varphi \setminus \{0\}$.
%We denote by $\leq _{tail}$ the natural total order on $tail_\varphi$, and $\leq$ the natural
%order on $\mX$.
A \emph{reset} on $\mX$ consists in enforcing any value of $\mX $ to $-\alpha$.  
%
%In this section, we briefly state some assertions around the asynchronous unison problem and its solutions.  
%
%
%\paragraph{ $WU$ predicate, and Path Delay }
%and Precedence Relation
%
We assume that each process $p$ maintains a clock register $p.r$ using an incrementing system $(\mX,\varphi)$. 
Let $\gamma $ be a system configuration, we define the predicate $WU$ as follows:
$
WU(\gamma ) \stackrel{\mathrm{def}}{\equiv} \forall p \in V,\forall q\in \mN_p:(p.r \in ring_{\varphi })\wedge (  | p.r - q.r | \leq 1)  \text{ in } \gamma
$.
In the remainder, we will abuse notation, referring to the corresponding set of 
configurations simply by $WU$. %Now, let us recall the notion of path delay and some of its properties established in \cite{BPV04b}.  

%\paragraph{Intrinsic Path Delay \cite{BPV04b}.}
%Let $\gamma$ a  configurations i
In $WU$, the clock values of neighboring processes are locally comparable.  In the sequel of the paper,
we need the three following definitions:
\BEGLIST
%\begin{enumerate}
\item [\textbf{Delay}.]  \label{def:delay} The delay of a path $\mu = p_0p_1\ldots p_k$, denoted by
$\delta _\mu$, is the local variation of the sequence 
$p_0.r, p_1.r,\ldots, p_k.r$, \ie $\delta _\mu = \sum \limits_{i=0}^{k-1}\left( p_{i+1}.r \ominus _l p_i.r\right)$ if 
$k > 0$, $0$ otherwise ($k=0$).
\item  [\textbf{Intrinsic Delay}.] \label{def:intrinsic}
The delay between two processes $p$ and $q$ is \emph{intrinsic} if it is independent on the choice
of the path from $p$ to $q$.  The delay is \emph{intrinsic} iff it is \emph{intrinsic} for every 
$p$ and $q$ in $V$. In this case, and at time $t$, the intrinsic delay between $p$ and $q$ is denoted by $\delta_{(p,q)}$.
\item [\textbf{WU$_0$}.] The predicate $ WU_0$ is true for a system configuration $\gamma$ iff $\gamma$ satisfies $WU$ 
and the delay is intrinsic in $\gamma$.  
%\end{enumerate}
\ENDLIST

%
%\subsection{Distributed Unison}
%
%\emph{Self-Stabilizing Asynchronous Unison} is a relaxed 
%\emph{self-stabilizing Barrier Synchronization} in the following meaning: 
%the clocks are in phase if the values of two neighboring processes  differ by no more than $1$, 
%and the clock value of each process is incremented by $1$ infinitely often. 
%Self-stabilizing Unison was introduced by \cite{CFG92}.  Let WU be the set of states where two neighboring processes  differ by no more than 1. It is easy to show that any guarded incrementing action,  for which WU is closed, is equivalent to the guarded action~(\ref{eq:guarde}), next page. Unfortunately there is a possibility of deadlock if the size of the clock is too short.  Moreover, if we want to stabilize Unison with a reasonable time complexity, we must use an incrementing system. 
%A little algebraic framework, and some vocabulary  are necessary. The vocabulary will be used in the definition of the  algorithm~\ref{algo:SSWS}. Following \cite{BPV04b}, we briefly introduce the important notion of \emph{intrinsic delay}, and the notion of \emph{cyclomatic characteristic}. Then, we define Unison problem and we give  the state of art about it.

\paragraph{Unison Definition.}

Assume that each process $p$ maintains a register $p.r \in \chi$. 
The self-stabilizing asynchronous unison problem, or simply the \emph{unison} problem, consists in the 
design of a protocol so that the following properties are true in every execution~\cite{BPV05}:  
\begin{list}{}{\partopsep -2pt \parsep -2pt \listparindent .3in \labelwidth .3in}
\item[\textbf{Safety }: ] $WU$ is closed.
\item[\textbf{Synchronization}: ] 
In $WU$, a process can increment its clock $p.r$ only if the value of $p.r$ is lower than or equal to
the clock value of all its neighbors. 
%, \ie $\forall q \in \mN_p,\ p.r \leq_l r_q$; 
\item[\textbf{No Lockout (Liveness)}: ]
In $WU$, every process $p$ increments its clock $p.r$ infinitely often. 
 \item[\textbf{Convergence}: ] 
$\Gamma \triangleright WU$. 
\end{list}

The following guarded action solves the \emph{synchronization property} and the \emph{safety}:
\begin{center}
$
%\begin{equation}
%\label{eq:guarde}
\forall q \in \mN_p:\ (q.r=p.r)\vee (q.r =\varphi (p.r))\longrightarrow p.r := \varphi(p.r)
$%\end{equation}
\end{center}
%Remark that If $K \geq 4$ and the safety is satisfied, then the synchronization is satisfied. 
 The predicate $WU_0$ is closed for any execution of this guarded action.  Moreover,
for any execution starting from a configuration in $WU_0$, the \emph{no lockout property} is
guaranteed. Generally, this property is not guaranteed in $WU$.

%\begin{lemma}
%If a uniform distributed protocol $\mP$ solves the asynchronous distributed unison problem, then any action in $WU$ is functionally equivalent to 
%the called Normal  Guarded Instruction:
%$$
% \forall q \in \mN_p:\ (q.r=p.r)\vee (q.r =\varphi (p.r)) \longrightarrow p.r := \varphi(p.r);
%$$
%\end{lemma}
%Conversely, \cite{BPV04b}   shows that if $K>C_G$ then the above guarded action solves the asynchronous unison problem in $WU$. 
%If $K\leq C_G$, there is a possibility of deadlocks~\cite{BPV04b}.

%A few general schemes to self-stabilizing the non-stabilizing protocols have been proposed. 
%The first self-stabilizing asynchronous unison was introduced in~\cite{CFG92}.  The deterministic protocol 
% proposed needs $K \geq n^2$. The stabilization time complexity is in $O(nD)$.
%The second solution is proposed in \cite{BPV04b}.  The authors show that  if  $K$ is
%greater than $C_G$  then $WU=WU_0$ and the \emph{no lockout property} is
%guaranteed in $WU$. 
%(see Definition~\ref{def:CG}).
%They present a solution using an incrementing system. The protocol is self-stabilizing if   $\alpha \geq T_G-2$, where $T_G$ is the length of 
%the longest chordless cycle ($2$ in tree networks).
%One can notice that $C_G$ and $T_G$ are bounded by $n$.  So, even if $C_G$ and $T_G$ are unknown, 
%we can choose $K \geq n+1$ and $\alpha=n$. Its self-stabilizing time complexity is in $O(n)$.
%In \cite{BPV06}, the authors present the Protocol $WU\_Min$, which is 
%self-stabilizing to asynchronous unison in at most $D$ rounds in trees.

\subsubsection{Protocol }

\paragraph{Variable and algorithm description.} 
The protocol is shown in Algorithm~\ref{algo:SSWS}.
For each process $p$, let $V(p,\varrho)$ be the set of processes which are cooperating (or conflicting) with $p$.  
Each process $q \in V(p,\varrho)$ is at most $k$-hops away from $p$ --- $d(p,q) \leq \varrho$.
Let $(\chi,\varphi)$ be an incrementing system, such that $\chi = \left\{ -\alpha ,..,0,1,...,\varrho K-1\right\}$.
In~\cite{BPV04b}, it is shown that:
\begin{enumerate}
\item $\alpha$  greater than or equal to $T_G$  ensures the convergence property of the unison, 
where $T_G$ is the \emph{size of the greatest hole} of $G$, \ie the length of the longest chordless cycle of $G$ if $G$ contains
cycle, $2$ otherwise ($G$ is acyclic);
\item $\varrho K$  greater than $C_G$  ensures the liveness property of the unison in $WU_0$, 
where $C_G$ is the \emph{cyclomatic characteristic} of $G$, \ie the smallest length of the longest 
cycle in the set of all the cycle basis of $G$. 
\end{enumerate}
Note that $T_G$ is upper bounded by $n$ and $C_G$ is upper bounded by $min(n,2D)$.
We assume that the above two conditions are satisfied.

%So, if $\varrho \geq D$, Algorithm~\ref{algo:SSWS} implements a wave.  

%\paragraph{Cyclomatic Characteristic $C_G$ \cite{BPV04b}.}
%\label{def:CG}
%The \emph{cyclomatic characteristic} $C_G$ is equal to $2$ or the length the lowest 
%$\lambda ( \Lambda ) $ in the set of cycle bases whether $G$ is acyclic or not, respectively. 
%Let $\lambda ( \Lambda ) $ be the length of the longest cycle of any cycle base $\Lambda$  of $G$. 
%any cycle basis of \Lamba$ of the longest 
%cycle in $\Lambda $ is denote
%
%
%d $\lambda (\Lambda)$ if $G$ is an acyclic graph, . 
%Otherwise $G$ contains cycles, then  let $\Lambda$ be a cycle basis.  The length of the longest 
%cycle in $\Lambda $ is denoted $\lambda (\Lambda)$.
%The cyclomatic characteristic of $G$, $C_G$, is equal to the lowest 
%$\lambda ( \Lambda ) $ among cycle bases.
%
%
%
%
%If we want to program a Barrier Synchronization, we must take $\varrho \geq D$. If $K\geq 3$ then
%the inequality $K\varrho >C_{G}$ holds. In the remainder, we suppose that the inequality $K\varrho >C_{G}$  holds.

\begin{algorithm}
\begin{footnotesize}
\noindent
{\bf Constant and variable}:\\
\hspace*{0.3cm}
$\mN_p$: the set of neighbors of process $p$; % \\
%\noindent
%\hspace*{0.3cm}
$p.r \in \chi $;\\
\noindent
{\bf Boolean Functions}:\\
\noindent
\hspace*{0.3cm}
\begin{tabular}{@{}lcl}
$ConvergenceStep_p$&$\equiv$ &$p.r \in tail_{\varphi }^{*}\wedge (\forall q\in \mN_p:(q.r \in tail_{\varphi })\wedge (p.r\leq _{tail_{\varphi }}q.r))$;\\
$LocallyCorrect_p$ &$\equiv$ & $p.r \in ring_{\varphi}\wedge (\forall q\in \mN_p,q.r \in ring_{\varphi}\wedge ( \left( p.r=q.r\right) \vee \left( p.r=\varphi \left(
q.r\right) \right) \vee \left( \varphi \left( p.r\right) =q.r\right) ))$;\\
$NormalStep_p$ &$\equiv$ & $p.r \in ring_{\varphi}\wedge (\forall q \in \mN_p:\ (p.r=q.r)\vee (q.r =\varphi (p.r)))$;\\
$ResetInit_p $ &$\equiv$ &
      $\neg LocallyCorrect_p\wedge (p.r \not\in tail_\varphi)$;\\
\end{tabular}\\
{\bf Actions}:\\
\noindent
\hspace*{0.3cm}
\begin{tabular}{@{}rlcl}
 %R1
   $NA:$ &  $NormalStep_p$ & $\longrightarrow$ & $<<\mbox{ CS 1}>>$ ; \\
   &&& if $p.r\equiv \varrho -1 [\varrho ]$ then $ <<\mbox{ CS 2}>>$ ;\\
   &&&  $p.r := \varphi(p.r)$;\\

%R2
   $CA:$ & %\ 
   $ConvergenceStep_p$ & $\longrightarrow$ &
   $p.r := \varphi(p.r)$;\\
%R3
   $RA:$ & %\ 
   $ResetInit_p$ & $\longrightarrow$ &
   $p.r := \alpha$ (reset);\\

\end{tabular}
\end{footnotesize}
\caption{($SS-WS$) The Self-Stabilizing Wave Stream  for $p$}
\label{algo:SSWS}
\end{algorithm}

\paragraph{Analysis of Algorithm~\ref{algo:SSWS} in $WU_0$.}
\label{sec:Unison_analysis}

%\paragraph{Precedence Relationship.}
%\label{rem:delay}
%It follows from the definition of  $C_G$ that $C_G \leq 2D$.
%\cite{BPV04b} shows that if $K> C_G$ the delay is intrinsic in $WU$. 
By definition of $WU_0$, the delay is intrinsic---refer to Subsection~\ref{sub:unison}. 
It defines a total preordering  on the processes in $V$, so called \emph{precedence relationship}. 
%This relationship depends on the state $\gamma \in WU_0$.
given a configuration in $WU_0$, the absolute value of the delay between two processes  
$p$ and $q$ is equal to or less than the distance $d(p,q)$ in the network. 
%
%Let us define Predicate $WU_0$ which is true for a system configuration $\gamma$ iff $\gamma$ satisfy $WU$ 
%and the delay is intrinsic in $\gamma$.  
%Clearly, the residuals are congruent to $0$ modulo $K$. 
%The delay being intrinsic iff it is equal to $0$ on every cycle, by linearity, 
%a path delay is intrinsic iff it is equal to $0$ on a cycle basis \cite{B89}.  
%We borrow the following definition from \cite{BPV04b}:
%
%
%
%\paragraph{Lifting Construction.} 

We will now prove that Algorithm~\ref{algo:SSWS} provides a $\varrho$-wavelet scheme.  We will develop a proof technique 
called \emph{lifting}.  The idea behind this term is to interpret any possible configuration in $WU_0$ 
by another such that the register values are in $\Bbb{N}$, the set of the positive integers.  In this way, 
the precedence relationship becomes the natural order on $\Bbb{N}$. It is possible because delay is intrinsic.

Consider $\gamma _0 \gamma _1...$ be a maximal execution starting in $WU_0$.  
%We show  how to unwind of the register $r$--The lifting .  Let $\gamma _{t_{0}}\gamma _{t_{0}+1}....$ be an  infinite execution 
%starting in $WU_0$. 
Let $p_{0}$ be a minimal process, according to the \emph{precedence relation}% -- see Remark~\ref{rem:delay} -- 
in $\gamma _{0}$. Let $\bot _0 = p_{0}.r$ at time $0$.
Denote the value of a register $r$ of a process $p$ in the state $\gamma_t$ by $p^t.r$. %$i$ is the moment of the state $\gamma_i$.
Similarly, $\delta^t_{(p,q)}$ denotes the delay between $p$ and $q$ in $\gamma_t$. 

For each process $p\in V$, %we unwind the register $p.r$ in the following manner. We
we associate the virtual register $\widetilde{p.r}$. For the state $\gamma _{0}$,
we initiate this virtual register by  the instruction 
$\widetilde{p.r}:=\bot _{0}+\delta _{(p_{0},p)}^{0}$. During the execution, for each  transition $\gamma
_{t}\mapsto \gamma _{t+1}$ the instruction $\widetilde{p.r}:=\widetilde{p.r}+1$ holds 
if and only if $p.r:=\overline{p.r+1}$ holds during the same transition.
Denote by $t_{p,k}$ the smallest time such that $\widetilde{p.r}:=k$.  Since the delay is
bounded by $D$, if $k\geq \bot _{0}+ D$, then $t_{p,k}$ is well defined and the cut  $C_{k}=\{ (p,t_{p,k}),p\in V\}$ is well defined on the network.
 
We now need to prove that for every $k \geq \bot_0 +D$, the cuts $C_{k}$ are coherent.  We first claim the following obvious lemma:

%Mainly, there are two possible assumptions, the first one is to suppose that at the beginning of the new computation all the network is correct, 
%and that no transient fault occurs during the calculation. 
%It is the notion developed in general distributed  algorithms theory. 

\begin{lemma}
\label{lem:coherent}
If $\left( p,t\right) \leadsto $ $\left( q,t^{\prime }\right) $ then: 
$
\widetilde{q^{t^{\prime }}.r}\in \left\{ \widetilde{p^{t}.r},\widetilde{p^{t}.r}+1\right\} 
$
 
Inductively, if $\left( q_{0},t_{0}\right) \leadsto $
$\left( q_{1},t_{1}\right) \leadsto $ $\left( q_{2},t_{2}\right) ...\leadsto 
$ $\left( q_{i},t_{i}\right) $ then: 
$
\widetilde{q_{i}^{t_{i}}.r}\in \left\{ 
\widetilde{q_{0}^{t_{0}}.r},...,\widetilde{q_{0}^{t_{0}}.r}+i\right\}
$
\end{lemma}

From the Lemma~\ref{lem:coherent},  
if $\left( q,t\right) \preceq \left( p,t_{p,k}\right) $ then $\left( q,t\right)
\preceq \left( q,t_{q,k}\right) $. It follows:

\begin{lemma}
For every $k\geq \bot _{0} +D$ the cut $C_{k}$ is coherent.
\end{lemma}

%Let $\Sigma _{p}^{\varrho }$ be the set of simple walks of length less than
%or equal to $\varrho $, ending to $p$. 

\begin{lemma}
\label{lem:cover}
Let $k\geq \bot_0 + D$. If $\left( p,t\right) $ is an event in the interval  
$\left[ C_{k},\rightarrow \right[ $, then $V(p,\widetilde{p^{t}.r}-k)\subset
Cover(p,t)$.
\end{lemma}

\begin{proof}
The statement holds for the initial events of $\left[ C_{k},\rightarrow
\right[ $. Let $\mathcal{A}$ be the set of events $(p,t)$ in  $\left[
C_{k},\rightarrow \right[ $ such that $V(p,\widetilde{p^{t}.r}-k)\subset
Cover(p,t)$ does not hold.  We assume that $\mathcal{A}$ is not empty, let $(q,\tau )$ a minimal event in $\mathcal{A}$ according to $\preceq $. 
Let $\varrho =\widetilde{q^{\tau }.r}-k$, and let $p_{1}\in V(q,\varrho )$. 
If $p_{1}=q$ then $p_{1}\in Cover(q,\tau )$, otherwise there exists $q_{1}\in 
\mathcal{N}_{q}$ such that $p_{1}\in V(q_{1},\varrho -1)$. $(q,\tau )$ is not
an initial event, so $q_{1}\in \mathcal{N}_{q}^{\tau }$ and there exists $%
\tau _{q_{1}}$ such that $\left( q_{1},\tau _{q_{1}}\right) \leadsto $ $%
\left( q,\tau \right)$.  By the minimality of $(q,\tau )$, 
$V(q_{1},\varrho -1) \subset Cover(q_{1},\tau _{q_{1}})$ holds.  So,  
$V(q_{1},\varrho -1)\subset Cover(q,\tau )$. Thus, $p_{1}\in Cover(q,\tau )$.
Therefore, $(q,\tau )$ is not in $\mathcal{A}$. Thus  $\mathcal{A=}\emptyset $, 
and the lemma is proved.
\end{proof}

As a corollary of Lemma~\ref{lem:cover}, the following result holds : 

\begin{theorem}
\label{th:unison_behavior}
Let $k\geq \bot_0 +D$ and $\varrho $ be a positive integer. Then, 
$\left[C_{k},C_{k+\varrho }\right] $, with $C_{k+\varrho }$ as the set of decide events, 
is a $\varrho$-wavelet.
%and a \emph{wave}
%if $\varrho \geq D$. 
\end{theorem}

\subsection{Infimum Computation}
\label{ssec:infimum_calcul}

\paragraph{Problem definition.}
In \cite{T88,Tel94}, the author introduces the infimum
operators. An infimum $\oplus $ over a set $\Bbb{S}$, is an associative,
commutative and idempotent (i.e. $x\oplus x=x$) binary operator. 
If $P=\left\{ a_{1},a_{2},...,a_{r}\right\} $ is a finite part of $\Bbb(S)$ and $a_0 \in \Bbb(S)$  then, from the associativity,
$\oplus P$ means $a_{1}\oplus a_{2}\oplus ...\oplus a_{r}$.  So, $a_0 \oplus P = a_0 \oplus a_{1}\oplus a_{2}\oplus ...\oplus a_{r}$. 
Such an operator defines a partial order relation $\leq _{\oplus }$ over $\Bbb{S}$,
by $x$ $\leq _{\oplus }y$ if and only if $x\oplus y=x$.\ We assume that $\Bbb{S}$
has a greatest element $e_{\oplus }$, such that $x$ $\leq _{\oplus }e_{\oplus }$
for every $x\in \Bbb{S } .$ Hence $\left( 
\Bbb{S } ,\oplus \right) $ is an Abelian
idempotent semi-group with $e_{\oplus }$ as identity element for $\oplus $.

\begin{theorem}[\cite{T88,Tel94}]
\label{th:infimum}
A wave can be used to compute an infimum.
\end{theorem}

%\begin{proof}
%%
%We claim that at the end of each event $(p,t)$ the register $p.res$ is equal to $\bigoplus \left\{ q.v_0,q\in Cover((p,t))\right\} $. 
%Let $\mathcal{A}$ be the set of events $(p,t)$ such that  $p.res \neq \bigoplus \left\{ q.v_0,q\in Cover((p,t))\right\} $.  If $\mathcal{A}$ is empty, the proof is finished. Note that the %minimal events in $\left[ C_{1},C_{2}\right] $ are not in $\mathcal{A}$. Suppose that%
%$\mathcal{A}$ is not empty.  Let $(p,t)$ a minimal event of $\mathcal{A}$  according to the relation $ \preceq $. At time $t$, $p.res:=p.v_0\bigoplus \left\{ q.res,q\in \mathcal{N}_{p}^{t}\right\} $. If $q\in \mathcal{N}_{p}^{t}$, we denote $t_{q}$ the time such that $\left(q,t_{q}\right) \leadsto \left( p,t\right) $, 
%then $Cover(p,t)=\left\{ p\right\} \bigcup\limits_{q\in \mathcal{N}_{p}^{t},}Cover\left(
%q,t_{q}\right) $. Because of the minimality of the event $(p,t)$, we deduce that for each $q \in \mathcal{N}_{p}^{t}$,   the equality $q.res = \bigoplus \left\{ s.v_0.,s\in Cover((q,t))\right\} $ holds at time $t_q$. It results that $p.res = \bigoplus \left\{ q.v_0.,q\in Cover((p,t))\right\} $ and $\mathcal{A}$ is empty.
%We know that for any decide event $(p,t)$ of the sequence $\left[ C_{1},C_{2}\right] $,  $Cover(p,t)=V$. Thus, following the result above, 
 %$p.v=\bigoplus \left\{ q.v_0.,q\in V \right\}$ at time $t$.

%\end{proof}

%\subsection{Unison as a self-stabilizing wavelet stream algorithm }
%
%\label{sec:Bar_syn}
\paragraph{Self-Stabilizing Infimum Computation in a $\varrho$-ball.}

%\subsection{Self-stabilizing computation of an infimum at distance  $\varrho$}
%\label{subsub:infimum_cal}
In order to  add a initializing step, we assume $\delta = \varrho +1$. 
We consider the following problem: at time $C_{U\delta }$ each register $p.v_0$ is initialized during the critical
section $<<CR2>>$, precisely when the register 
$p.r$ takes the value $UC\delta  $. At the end of each phase $\Phi_U=
$ $\left[ C_{U\delta },C_{U\delta  +\delta -1}\right] $ , each process $p$
needs to known the infimum of the registers $q.v_0$ of every
$q$ in $V\left( p,\varrho \right)$.

To reach the objective, we define for each process $p$ two added registers 
$p.v_{1}$ and $p.v_{2}$. These two registers are initialized  at the date $C_{U\delta }$ during the critical section $<<CR2>>$, by the value $p.v_{0}$.

For $k \in \left\{ 1,2,...,\varrho \right\} $, at the date $C_{U\delta + k }$, 
the action $<<CS1>>$ is defined by:\newline
\noindent
$
p.v_{1}:=p.v_{2};\ p.v_{2}:=p.v_{0}\bigoplus \left\{ q.v_{\omega  \left(
q\right) },q\in \mathcal{N}_{p}\right\} 
$,
%\end{center}
with,  if $q.r=p.r$ then  $\omega  \left( q\right) =2$, and if $q.r=p.r+1$  
then $\omega  \left( q\right) =1$.

\begin{proposition}
For $p \in V$ and $k \in \left\{ 1,...,\varrho \right\} $, at the date
$C_{U\delta  + k}$, both equalities hold:\newline
\noindent 
$(1)$ $p.v_{1}=\bigoplus \left\{ q.v_{0},q\in V\left( p,k -1\right) \right\}$,
and $(2)$ $p.v_{2}=\bigoplus \left\{ q.v_{0},q\in V\left( p,k \right)
\right\} 
$. 
\end{proposition}

\begin{proof}

At $C_{U\delta  },$ any process $p$ satisfies $p.v_{1}=p.v_{0}$
and $p.v_{2}=p.v_{0}$, it is the initializing step. Let $\mathcal{A}$ the set
of events in $\Phi_U=\left[ C_{U\delta  +1},C_{U\delta  +\varrho }\right] $, for
which the proposition is not true. 

Assume by contradiction that $\mathcal{A}$ is not
empty. Let $(p,t)$ a minimal event in $\mathcal{A}$. Let $k \in \left\{
1,2,...,\varrho \right\} $ such that $(p,t)$ $\in C_{U\delta  + k }$.
There exists $t_{0}$ such that $\left( p,t_{0}\right) \in \Phi_U $ and $\left( p,t_{0}\right) \rightsquigarrow (p,t)$. 
We have $p^{t}.v_{1}=p^{t_{0}}.v_{2}=$ and $p^{t_{0}}.v_{2}=\bigoplus
\left\{ q.v_{0},q\in V\left( p,k -1\right) \right\} $. This equality is
true even if $k =1$.
Now, $p^{t}.v_{2}=p.v_{0}\bigoplus \left\{ q^{t_{q}}.v_{\varphi \left(
q\right) },q\in \mathcal{N}_{p}\right\} $. From the minimality of the event $%
\left( p,t\right) $, the events $\left( q,t_{q}\right) $, where $t_{q}<t$, 
are not in $\mathcal{A}$ and are in $\left[ C_{U\varrho },C_{U\varrho +\varrho
-1}\right] $. 
So, $p.v_{0}\bigoplus \left\{ q^{t_{q}}.v_{\varphi \left(
q\right) },q\in \mathcal{N}_{p}\right\} =\bigoplus \left\{ q.v_{0},q\in
V\left( p,k \right) \right\} $. We obtain a contradiction. 
%We deduce that $\mathcal{A}$ is empty and the proposition follows.
\end{proof}

As a corollary, we obtain the expected theorem:

\begin{theorem} %[\cite{Bou06}]
On the cut $C_{U\delta  +\varrho } $, 
 $p.v_2$ contains the infimum of the registers $q.v_0 $ in the $\varrho-$ball centered in $p$, according to the phase $U=\left[ C_{U\delta  },C_{U\delta  +\varrho }\right] $.
\end{theorem}

\section{Applications}
\label{sec:distance_scheme}

In this part, we show how to synchronize a self-stabilizing layer clock. The main clock defines a wavelet stream. 
Using the wavelet stream, we design with the slave clock a barrier synchronization at distance $\rho $. 
We then show how to use this layer clock to tackle efficiently many local synchronization problems at distance $\rho$.

\subsection{Self-stabilizing Layer Clock}
\label{subsec:layer_clock}

The idea is to manage the $\varrho-$wavelet stream. A clock organizes
this stream.  The wavelets are used to compute concurrently local infimum on
each  $\varrho$-ball. For each process $p$, once the infimum computed, a second
clock defines a delay notion on the network. This delay is a total
preordering useful to schedule the critical section enter of each process. 

%It is why, to schedule the
%critical section enter, each process  must have an identity.
%Combining the ordering defined by the identities and the preordering defined
%by the delay of the second clock, we define a dynamic ordering on the
%processes. In fact it is only necessary to define an  preordering on the
%processes in each $\varrho -$balls. 

%So it should be sufficient that
%identities of processes are different at distance $\varrho $.
 
More formally, we define two clocks, the first clock $C_1$ (the master clock) and a second clock $C_2$ (the slave clock).  
The incrementing systems are respectively $\left( \chi_{1},\varphi _{1}\right) $ 
and    $\left( \chi _{2},\varphi _{2}\right) $. The behavior of the slave clock  is scheduled by
the first clock and a predicate $cond$. The predicate $cond$ depends of the problem solved. 
To distinguish the two clocks, all the registers are subscripted by $1$ or $2$ respectively for the
master clock and the slave clock. The predicates are superscripted by $1$ or $2$.  For instance, the register of the master
clock is denoted by $r_{1}$ and the register of the slave clock is denoted
by $r_{2}$, and the predicate $NormalStep_{p}^{1}$ is defined on the register 
$r_{1}$ of the process $p$.  The predicate $NormalStep_{p}^{2}$ is defined on
the register $r_{2}$ of the process $p$.  We define in the same way $WU_{1}$, $WU_{2}$,
and $WU=WU_{1}$ $\cap $ $WU_{2}$.
The stabilization of the layer clock is to guarantee $\Gamma \rhd WU.$

When the system is stabilized,  the schedule of the slave clock of a process $p$ is defined in $<<CS2>>$ by 
\begin{center}
%\begin{equation}
$
NormalStep_{p}^{2}\wedge cond \rightarrow\ <<CS2>>;\ p.r_{2}:=\varphi _{2}\left(
p.r_{2}\right) 
$
%\end{equation}
\end{center}
Predicate $cond$ is independent of the register $p.r_1$. It is this predicate which expresses  the distance $\varrho$  synchronization problem solved, 
while the procedures \emph{Initialization} and \emph{Computation} are 
scheduled by the wavelet and  provide a preprocessing used by $cond$.
The procedures \emph{Initialization} and \emph{Computation} depends on the problem to be solved to the layer clock.  
We give some instances for different problems later.  

We define the predicate $NormalStep_p\equiv NormalStep_p^1 \wedge LocallyCorrect_p^2$.

\begin{algorithm}
\begin{footnotesize}
\noindent
{\bf Constant and variable}:\\
\hspace*{0.3cm}
$\mN_p$: the set of neighbors of the process $p$;  \\
\noindent
%\hspace*{0.3cm}
$p.r_1 \in \chi_1 $;
$p.r_2 \in \chi_2 $;\\
\noindent
{\bf Boolean Functions}:\\
\noindent
{\bf For clock $i \in \{ 1,2 \}$}:
\noindent

\hspace*{0.3cm}
\begin{tabular}{@{}lcl}
$ConvergenceStep_p^i$&$\equiv$ &$p.r_i \in tail_{\varphi_i }^{*}\wedge (\forall q\in \mN_p:(q.r_i \in tail_{\varphi_i })\wedge (p.r_i\leq _{tail_{\varphi_i }}q.r_i))$;\\
$LocallyCorrect_p^i$ &$\equiv$ & $p.r_i \in ring_{\varphi_i}\wedge $\\
&& $\forall q\in \mN_p,q.r \in ring_{\varphi_i}\wedge ( \left( p.r_i=q.r_i\right) \vee \left( p.r_i=\varphi_i \left(
q.r_i\right) \right) \vee \left( \varphi_i \left( p.r_i\right) =q.r_i\right) )$;\\
$NormalStep_p^i$ &$\equiv$ & $p.r_i \in ring_{\varphi_i}\wedge (\forall q \in \mN_p:\ (p.r_i=q.r_i)\vee (q.r_i =\varphi (p.r_i)))$;\\
$ResetInit_p^i $ &$\equiv$ &
$\neg LocallyCorrect_p^i\wedge (p.r_i \not\in tail_{\varphi_i})$;\\
\end{tabular}\\
\noindent
\bf{ Common predicate}:\\
\noindent
\hspace*{0.3cm}
\begin{tabular}{@{}lcl}
$NormalStep_p $&$\equiv$&$ NormalStep_p^1 \wedge LocallyCorrect_p^2$;\\
\end{tabular}\\
\noindent
{\bf Actions}:\\
\noindent
\hspace*{0.3cm}
%\noindent
\begin{tabular}{@{}rlcll}
 %R1
   $NA:$ &  $NormalStep_p$ & $\longrightarrow$ & if $p.r_1\equiv \varrho -1 [\varrho ]$ then &\\
      &&&  \  \ \ \text{  } Begin & \\   
     &&&  \  \ \ \text{  } $NormalStep_{p}^{2}\wedge cond\rightarrow $ & $<<\mbox{ CS 2}>> $  ;\\
     &&&                                                       & if $cond_1$ then $p.r_{2}:=\varphi _{2}\left(p.r_{2}\right)$  \\
     &&&\  \ \ \text{  } $Initialization$ ; & \\
      &&&  \  \ \ \text{  } End & \\ 
      &&& else $Computation$; & \\
    &&& $p.r_1 := \varphi_1(p.r_1)$; &\\
\end{tabular}\\
\noindent
{\bf For clock $i \in \{ 1,2 \}$}:\\
\noindent
\hspace*{0.3cm}
\begin{tabular}{@{}rlcl}  
%R2
   $CA_i:$ & %\ 
   $ConvergenceStep_p^i$ & $\longrightarrow$ &
   $p.r_i := \varphi_i(p.r_i)$;\\
   
%R3
   $RA_i:$ & %\ 
   $ResetInit_p^i$ & $\longrightarrow$ &
   $p.r_i := \alpha_i$ (reset);\\

\end{tabular}
\end{footnotesize}
\caption{($SS-DC$) Self-stabilizing Layer Clock}
\label{algo:layer_clock}
\end{algorithm}

Due to the lack of space, the proofs of Proposition~\ref{cor:WUstab} and~\ref{prop:nos} are left in the appendix.

\begin{proposition}
\label{cor:WUstab}
The layer clock stabilizes to $WU$.
\end{proposition}

\begin{proposition}[No starvation]
\label{prop:nos}
Once stabilized, the clock \emph{C}$_{1}$ increments infinitely often.
\end{proposition}

\subsection{Local Comparison in a $\varrho$-ball}

In the network, when the layer clocks are  stabilized, the delay between two processes according to the slave clocks   defines a total preordering  on the processes. 
Unfortunately this delay is a global notion. The problem is to
find a condition such that for two processes in the same $\varrho$-ball, it is
possible  to calculate directly the delay with only the knowledge of the $r_{2}$ register values, and so to decide which process precedes the other
according to the delay. 
To organize comparison between two processes lying  in a same ball of radius
equal to $\varrho $, it is sufficient to be able to compare the slave clock
registers  of any two  processes at distance less than or equal to $2\varrho $. By
this way,  in each $\varrho$-ball $B$, we will be able to define a total
preordering among the processes in $B$ by comparison of the values of the
registers $r_{2}$ of the processes. Of course we want that this total
preordering is the same than the preordering defined by the delay.
In order to reach this objective, we extend the locally comparability defined at one hop (refer to Section~{sub:unison}) 
to the distance $2\varrho$.  For the clarity, we must be
more formal:
A local order on a set $\chi $ is an antisymmetric and reflexive binary
relation on $\chi$.
Let $\chi=\{0,...,K-1\}$ such that $K\geq 4\varrho +1$. 
Let $a$ and $b$ be two elements of $\chi $. Let us assume that $d_{K}\left(
a,b\right) \leq 2\varrho $. Let us define now a local ordering  $\leq
_{l}$ by : 
$a\leq _{l}b\Leftrightarrow _{def}0\leq \overline{b-a}\leq 2\varrho $.

\begin{lemma}
Let $p$ and $q$ be two processes satisfying $d\left( p,q\right) \leq
2\varrho $ . If $a=p.r_{2}$ and $b=q.r_{2}$ then the delay $\varrho _{p,q}$ is
equal to $\overline{b-a}$ if $0\leq \overline{b-a}\leq 2\varrho $ and is
equal to $-\overline{a-b}$ otherwise$.$
\end{lemma}

\begin{proof}
Since $d\left( p,q\right) \leq 2\varrho $, we have $\varrho _{p,q}\in
\left\{ -2\varrho ,...,2\varrho \right\} $.  Moreover, $\varrho
_{p,q}\equiv \overline{b-a}\left[ K\right] $.  Since, $K>4\varrho $, 
then $\varrho _{p,q}$ is equal to $\overline{b-a}$ if $0\leq \overline{b-a}\leq
2\varrho $, $-\overline{a-b}$ otherwise.
\end{proof}

From this lemma, we access to the delay in each $\varrho -$ball $B.$ So our
problem is solved. 
In the following section, we assume that $\varrho =\varrho +1$, $K_1=\varrho K$ with $K_{1}\geq C_{G}-1$ and $%
K_{2}\geq \max \left( 4\varrho +1,C_{G}-1\right) $. These assumptions ensure 
that the layer clocks are self-stabilizing, that the main clock is calibrate to defined a $\varrho $-wave stream , and that delay defined by the
slave clocks is computable at distance $2\varrho $ with the only knowledge of
the  slave clock registers $r_{2}$.

\subsection{$\varrho$-Local Resource Allocation}
\label{subsec:usedc}

\subsubsection{Problem Definitions}

The \emph{Resource Allocation} problem deals with resource sharing problems among
the processes.  The resource allocation allows processes to access resources,
(\ie their critical sections) concurrently, provided the resources are not
conflicting with each other.

\begin{definition}[Graph of Compatibility~\cite{CDP03}] 
Let $A$ be a set -- sometime named the resource set--,
let $R$ be a reflexive binary relationship on $A.$ We say that $R$ is the
compatibility relationship on $A$. If $\left( a,b\right) \in R$, then we say
that $a$ and $b$ are compatible. If $\left( a,b\right) \notin R$ then we say
that $a$ and $b$ are conflicting.
\end{definition}

The specification of the general resource allocation problem is defined as follows:
\BEGLIST
%\begin{description}
\item  [{\bf Safety}:] if a processor requests a resource in $A$ to enter in
critical section, then its request is eventually satisfied and it enters the
critical section.

\item  [{\bf Fairness}:] In every execution, if two processes execute their
critical section simultaneously, then both are using resources whose are
compatible.
%\end{description}
\ENDLIST

Most of the problem requiring coordination among process sharing some resources are 
particular instances of the graph of compatibility, and then, particular safety requirements.  
For instance, the following well-known problems are particular instances of the resource allocation problem:

\BEGLIST
%\begin{enumerate}
\item  [\textbf{Mutual exclusion}:] $A$ is the set of processes and $R=\left\{ \left(
a,a\right) ,a\in A\right\}.$ The safety condition is:
\emph{In every execution, no two processes execute their critical section
simultaneously}.

\item  [\textbf{Readers-Writer}:] $A$ is the set of 2-uples $\left\{ \left( p,r\right)
,p\in V,r\in \left\{ read,write\right\} \right\} $, and $R$ is defined by
the safety condition:
\emph{ In every execution, if two processes execute their critical section
simultaneously, then both are executing a read operation}.

\item  [\textbf{Group Mutual Exclusion}:] $R$ is a equivalence relationship over a set
of resources $A$. $R$ is defined by the safety condition:
\emph{In every execution, if two processes execute their critical section
simultaneously, then both are using resources in the same equivalence
class}.

%\end{enumerate}
\ENDLIST

We now generalize the above requirements by limiting their effect to the conflict processes 
which are at $\varrho$-hops away of any given process $p$.  Obviously, if $\varrho = D$,
then the problem comes down to the above requirements.  If 
$\varrho =1$, then the set of processes which are conflicting with a process $p$ is reduced 
to the neighboring processes of $p$.  The most popular of this problems is the \emph{dining philosopher} problem,
also called the \emph{Local Mutual Exclusion} (LME) problem.

\BEGLIST
%\begin{enumerate}
\item  [\textbf{$\varrho$-Local mutual exclusion}:]
 $A$ is the set of processes and $R=\left\{
\left( a,b\right) \in A^{2},d\left( a,b\right) >\varrho \text{ or }%
a=b\right\} $

\item  [\textbf{$\varrho$-Local Readers-Writer}:] $A$ is the set of 2-uples $\left\{ \left(
p,r\right) ,p\in V,r\in \left\{ read,write\right\} \right\} $, and $R$ is
defined by the safety condition:
\emph{  In every execution, if two $\varrho$-neighboring processes execute
their critical section simultaneously, then both are executing a read
operation}.

\item  [\textbf{$\varrho$-Local Group Mutual Exclusion}:] $R$ is a equivalence relationship over
a set of resources $A$ . $R$ is defined by the safety condition:
\emph{ In every execution, if two $\varrho$-neighboring processes execute
their critical section simultaneously, then both are using resources in the
same equivalence class}.

%\end{enumerate}
\ENDLIST

The $\varrho$ generic version called the \textbf{$\varrho$-Local Resource Allocation} ($\varrho$-LRA) is 
specified as follows: $R$ is a general relationship over a set
of resources $A$. The safety condition is:
\emph{  In every execution, if two $\varrho$-neighboring processes execute
their critical section simultaneously, then both are using resources whose
are compatible}.
%
%
%There are many concrete
%instances which motivate this generalization like the $\varrho$-local
%calculation \cite{GMM04}, or the interferences in sensors networks \cite{DNT06}. 
%It is clear that global mutual exclusion is a way to solve all this
%problems.  The question is to increase the concurrency among the processes.  
%In the sequel, we provide solutions to the above problems.  

%\subsection{k-out-of-l exclusion problem}

%\begin{definition}
%There are $l$ units of a shared resource, any process can request at most $k$
%resources ($1\leq k\leq l)$ units of the shared resource, and no unit can be
%allocated to more than one process at one time. (re; RAYNAL).
%\end{definition}

%The specification of the $\varrho -$local generalization  problem is:

%\begin{description}
%\item  (Safety): if a processor requests a resource in $A$ to enter in
%critical section, then its request is eventually satisfied and it enters the
%critical section.

%\item  (Fairness): In every execution, if two processes execute their
%critical section simultaneously, then both are using resources whose are
%compatible.
%\end{description}

\subsubsection{Self-Stabilizing Solutions}
\label{sub:SSLRA}

Due to the lack of space, the correctness proofs of this section are left in the appendix.

\paragraph{$\varrho$-LRA.}

Each process $p$ maintains three registers $p.v:\Sigma $ where $\Sigma $ is
any data type, and the registers $p.res_{1}:\chi _{2}$ and $p.res_{2}:\chi _{2}\times \Sigma $. The
content of $p.v$ is the asked resource. Let us assume that a total order $
\preceq $ is defined on $\Sigma $. 
We reach the two fields of the register $p.res_{i}$ by $p.res_{i}.r$ and $%
p.res_{i}.v$ respectively, with $i\in \left\{ 1,2\right\} $.  
We define a $\varrho -$local ordering on $\chi _{2}\times \Sigma $ by:
\begin{center}
$
\left( r,v\right) \blacktriangleleft (r^{\prime },v^{\prime
})\Leftrightarrow  (r<_{l}r^{\prime })  \text{ or } (r=r^{\prime }\text{ and }v\preceq v^{\prime })
$
\end{center}
Recall that $<_{l}$ is defined by the delay, which is a total preordering. This preordering being computable at distance $\varrho$.    
Define the associated $\varrho$-local infimum in the following way:
\begin{center}$
\left( r,v\right) \oplus (r^{\prime },v^{\prime })= \text{ if } \left( r,v\right) 
\blacktriangleleft (r^{\prime},v^{\prime }) \text{ then } \left( r,v\right) \text{ else } (r^{\prime },v^{\prime })
$
\end{center}

Define the macros \emph{Initialization} and \emph{Computation}, respectively by:
\begin{center}
$
\begin{array}{rcl}
\text{Initialization } & \equiv & p.res_{1}:=\left( p.r_{2},p.v\right) \text{; }
p.res_{2}:=\left( p.r_{2},p.v\right) \\
Computation & \equiv & p.res_{1}:=p.res_{2}\text{; }p.res_{2}:=\left(
p.r_{2},p.v\right) \bigoplus \left\{ q.res_{\omega  \left( q\right) },q\in 
\mathcal{N}_{p}\right\} 
\end{array}
$
\end{center}
with, if $q.r_{2}=p.r_{2}$ then $\omega  \left( q\right) =2$ and if $q.r_{2}=%
\overline{p.r_{2}+1}$ then $\omega  \left( q\right) =1$. 
The preprocessing of a local infimum designates a winner $q$ in the $\varrho -
$ball centered in $p$. It is important to see that this process is elected
by $p$, and perhaps it is not elected by all the processes in the  $\varrho -$%
ball centered in $q$.  
The condition $cond$ depends of the solved problem, it is the disjunction: $
\left( p.r_{2},p.v\right) =$ $p.res_{2}$ or  $cond_{2}$. Where $\left(
p.r_{2},p.v\right) =$ $p.res_{2}$ means that $p$ is elected in the  $\varrho -
$ball centered in $p$, the condition $cond_{2}$ is there to raise
concurrency, it depends of the solved problem.
If  $p$ is not elected by the $\varrho -$local infimum calculation and $%
cond_{2}$ is true, the incrementation of the register may to be  not 
wanted. The condition $cond_{1}$ of the incrementation is: $p.r_{2}=p.res_{2}.r$.

%APPENDIX

\paragraph{$\varrho$-LME.}

Assume that each process $p$ has an identifier denoted by $p.id$. The
value of $p.id$ is in a total ordered set $\Bbb{S}$, for instance the
integers. In fact, the identities must be only a $2\varrho $-distance
network coloring. That is to say that every node must be colored
such that two vertices lying at distance less than or equal to $2\varrho $
have not the same color.
We define for each process $p$ the register $p.v$ as $p.id$. 
The couple $\left( p.r_{2},p.id\right) $ is defined for each process $p$. The condition $cond_2$ is defined by $false$ and thus $cond_1\equiv \TRUE$.

%PROOFS IN THE APPENDIX

\begin{definition}
$\varrho$-LME has a fairness index of $k$, if in any computation, between
any two consecutive critical section execution of a process, any other
process can execute its critical section at most $k$ times. The time-service
of $\varrho$-LME is the maximal number of critical executions by other
processes between two successive executions of the critical section by any
process.
\end{definition}

Our $\varrho$-LME algorithm is a barrier synchronization at distance $\varrho$. We deduce:

\begin{proposition}
For the $\varrho$-LME algorithm, the fairness index is  equal to $\left\lceil \frac{D}{\varrho }\right\rceil $, and 
the service time is upper bounded by $\left\lceil \frac{n\left( n-1\right) }{\varrho }\right\rceil $.
\end{proposition}

From the definition of a phase~\cite{BPV04b}, and because $K_{1}$ is in $O\left( \varrho D\right) \subset O\left( D^{2}\right)$,  
$K_{2}$ is in $O\left( D\right)$, respectively. So, $K_{1}K_{2}$ is in $O\left(
D^{3}\right)$. Thus:	

\begin{proposition}
$(1)$ During one phase, the number of link-communications is equal to $2\left( \varrho
+1\right) \left| E\right| $, where $\left| E\right| $ is the number of edges
in the network.
$(2)$ The stabilization time complexity of $\varrho$-LME algorithm is in $O\left( n\right)$ rounds, and
$(3)$ The space complexity of $\varrho$-LME algorithm is in $O\left( \log D\right) $.
\end{proposition}

Our solution for the $\varrho$-LME problem provide a good technique to reduce the service time 
and the fairness index of $LME$. The price to pay is the increase of the communications between
processes.

\paragraph{$\varrho$-Group mutual exclusion.} 

Let $\Sigma $ be the set of resources. Let us assume 
that the preordering $\preceq $ is
defined on $\Sigma $ --an arbitrary ordering, for instance a priority ordering. The
binary relationship on $A$, defined by: $x\asymp y$ iff $x\preceq y$ $and$ $
y\preceq x,$ is an equivalence relationship. The equivalence classes are the
groups. The set of groups is the quotient $\frac{\Sigma }{\asymp }$. 
$p.v$ takes its values in $\frac{\Sigma }{\asymp }$. An other way is to say that $\preceq $ is an total ordering on the groups.
To raise concurrency, if $p$ asks a resource $a$ and if the elected process at distance $\varrho $ requests a resource in the same group, $p$  enters in critical section.
The predicate $cond$ is defined as: $\overline{p.v}= p.res.v$.  

Note that we assume that there is no identity on the processes.  However, we make the additional assumption 
that there is a total ordering on the resources. For instance, Local Mutual Exclusion problem is an instance of 
Group Mutual Exclusion where the resources are the processes.  So, there is an ordering on the processes, which equivalent 
to define process identities.

%PROOFS IN THE APPENDIX

\paragraph{$\varrho$-Readers-Writer.}

We assume that each process has an identity denoted by $p.id$. Each process has three possible requests: the process does not ask
anything, the process asks to \emph{read}, the
process asks to \emph{write}. This requests are
symbolized respectively by $N,R,W$. In order to be able
to compare two registers $r$ at distance $\varrho $, we assume that $K\geq
4\varrho +1$. For each $\varrho $-ball $B$, the local ordering $\leq _{l}$
defines a total preordering on the registers $r$ of processes in $B$. 
If a process $p$ asks $N$ or $R$ then the register $p.v$ is initialized by
the value $F$. If $p$ asks  $W$ then $p.v$ is initialized
by the value $W p.id) \ $.  
The ordering $\preceq $ on $\Sigma $ is defined by:

\[
v\preceq v^{\prime } \stackrel{\mathrm{def}}{\equiv} 
\begin{array}{l}
(v=F\text{ and }v^{\prime }=F)\text{ or }\left( v=WId\ \text{and }v^{\prime
}=F\right)  \\ 
\text{or }\left( v=W\ Id\text{ and }v^{\prime }=W\ Id^{\prime }\text{ with }%
Id\leq Id^{\prime }\right) 
\end{array}
\]

For each process $p$, the predicate $cond$ is defined by matching on $p.res$ as follows: 
%\singlespacing
%\small
\[
\begin{array}{ll}
% match\text{ }p.res\text{ with:} &\\
\left( r,F\right) \text{ when } r =p.r_{2}\text{ } & \rightarrow \TRUE \\
|\ \left( r, W id\right) \text{ when } r =p.r_2\text{ and } p\text{ requests } N & \rightarrow \TRUE \\
|\ \left( r, W id\right) \text{ when }r =p.r_2\text{ and }\left(p.r_{2},p.id\right) =\left( r,id\right) \text{ } &\rightarrow \TRUE \\
|\ \_ \_ & \rightarrow  \FALSE
\end{array}
\]
%\normalsize
%\text{ when } r = p.r_2\text{ }&\rightarrow true \\
%\left( r, W id\right) \text{ when } r =p.r_2\text{ and } p\text{ requests }N & \rightarrow true \\
%\left( r, W id\right) \text{ when }r =p.r_2\text{ and }\left(p.r_{2},p.id\right) =\left( r,id\right) \text{ } &\rightarrow true \\
%\_ \_ &\rightarrow  false

%\doublespacing

\section{Conclusion}
\label{sec:conclusion}
We presented a self-stabilizing algorithm to solve the $\varrho$-wavelet scheme in arbritrary anonymous networks. 
Wavelets deals with coordination among processes which are at most $\varrho$ hops away of each other.  
The proposed algorithm works under any (even unfair) daemon. 
Using the wavelet scheme, we described a self-stabilizing layer clocks protocol and 
showed that it provides an efficient device in the design of local coordination 
problems at distance $\varrho$, \ie $\varrho$-barrier synchronization and
$\varrho$-local resource allocation (LRA) such as $\varrho$-local mutual exclusion (LME), 
$\varrho$-group mutual exclusion (GME), and $\varrho$-Reader/Writers.  Some solutions to 
the $\varrho$-LRA problem (\eg $\varrho$-LME) allow 
to transform algorithms written assuming any $\varrho$-central daemon into algorithms working 
with any distributed daemon.  

\singlespacing

%\footnotesize

\begin{small}
\bibliographystyle{alpha}
\bibliography{unison}%PFC,pfc,causal}

\newcommand{\etalchar}[1]{$^{#1}$}
\begin{thebibliography}{GGH{\etalchar{+}}04}

\bibitem[BDPV99]{BDPV99b}
A~Bui, AK~Datta, F~Petit, and V~Villain.
\newblock State-optimal snap-stabilizing {PIF} in tree networks.
\newblock In {\em Proceedings of the Forth Workshop on Self-Stabilizing Systems
  (WSS'99)}, pages 78--85. IEEE Computer Society Press, 1999.

\bibitem[BPV04]{BPV04b}
C~Boulinier, F~Petit, and V~Villain.
\newblock When graph theory helps self-stabilization.
\newblock In {\em PODC '04: Proceedings of the twenty-third annual ACM
  symposium on Principles of distributed computing}, pages 150--159, 2004.

\bibitem[BPV05]{BPV05}
C~Boulinier, F~Petit, and V~Villain.
\newblock Synchronous vs. asynchronous unison.
\newblock In {\em 7th Symposium on Self-Stabilizing Systems (SSS'05), LNCS
  3764}, pages 18--32, 2005.

\bibitem[BPV06]{BPV06}
C~Boulinier, F~Petit, and V~Villain.
\newblock Toward a time-optimal odd phase clock unison in trees.
\newblock In Springer-Verlag, editor, {\em Eighth International Symposium on
  Stabilization, Safety, and Security of Distributed Systems (SSS'06)}, Lecture
  Notes in Computer Science, 2006.

\bibitem[CDP03]{CDP03}
S~Cantarell, AK~Datta, and F~Petit.
\newblock Self-stabilizing atomicity refinement allowing neighborhood
  concurrency.
\newblock In Springer-Verlag, editor, {\em DSN SSS'03 Workshop: 6th Symposium
  on Self-Stabilizing Systems (SSS '03)}, volume 2704 of {\em Lecture Notes in
  Computer Science}, pages 102--112, 2003.

\bibitem[CHP71]{CHP71}
P.J. Courtois, F.~Heymans, and D.L. Parnas.
\newblock Concurrent control with readers and writers.
\newblock {\em Communications of the Association of the Computing Machinery},
  14(10):667--668, 1971.

\bibitem[Dij65]{D65}
E.W. Dijkstra.
\newblock Solution to a problem in concurrent programming control.
\newblock {\em Communications of the Association of the Computing Machinery},
  8(9):569, 1965.

\bibitem[Dij68]{Dij68}
E~Dijkstra.
\newblock {\em Cooperating Sequential Processes}.
\newblock Academic Press, 1968.

\bibitem[Dij74]{D74}
EW~Dijkstra.
\newblock Self stabilizing systems in spite of distributed control.
\newblock {\em Communications of the Association of the Computing Machinery},
  17:643--644, 1974.

\bibitem[DIM97]{DIM97a}
S~Dolev, A~Israeli, and S~Moran.
\newblock Uniform dynamic self-stabilizing leader election.
\newblock {\em IEEE Transactions on Parallel and Distributed Systems},
  8(4):424--440, 1997.

\bibitem[DNT06]{DNT06}
P~Danturi, M~Nesterenko, and S~Tixeuil.
\newblock Self-stabilizing philosophers with generic conflicts.
\newblock In {\em 8th International Symposium on Stabilizing, Safety, and
  Security of Distributed Systems (SSS'06)}, 2006.

\bibitem[Dol00]{D00}
S~Dolev.
\newblock {\em Self-Stabilization}.
\newblock The MIT Press, 2000.

\bibitem[GGH{\etalchar{+}}04]{GGHK04}
M~Gairing, W~Goddard, S~T Hedetniemi, P~Kristiansen, and A~A McRae.
\newblock Distance-two information in sefl-stabilizing algorithms.
\newblock {\em Parallel Processing Letters}, 14(3-4):387--398, 2004.

\bibitem[GH99]{GH99}
M~Gouda and F~Haddix.
\newblock The alternator.
\newblock In {\em Proceedings of the Fourth Workshop on Self-Stabilizing
  Systems}, pages 48--53. IEEE Computer Society Press, 1999.

\bibitem[GHJT06]{GHJT06}
W.~Goddard, S.T. Hedetniemi, D.P. Jacobs, and V.~Trevisian.
\newblock Distance-$k$ information in self-stabilizing algorithms.
\newblock In {\em The 13th International Colloquium On Structural Information
  and Communication Complexity Proceedings (SIROCCO'06), LNCS 4056}, pages
  349--356. Springer, 2006.

\bibitem[GMM04]{GMM04}
E.~Godard, Y.~M\'etivier, and A.~Muscholl.
\newblock Characterizations of classes of graphs recognizable by local
  computations.
\newblock {\em Theory of Computing Systems}, 37:249--293, 2004.

\bibitem[HC92]{HC92}
ST~Huang and NS~Chen.
\newblock A self-stabilizing algorithm for constructing breadth-first trees.
\newblock {\em Information Processing Letters}, 41:109--117, 1992.

\bibitem[Joh97]{J97}
C~Johnen.
\newblock Memory-efficient self-stabilizing algorithm to construct {BFS}
  spanning trees.
\newblock In {\em Third Workshop on Self-Stabilizing Systems}, pages 125--140.
  Carleton University Press, 1997.

\bibitem[Jou00]{Jou00}
Yuh-Jzer Joung.
\newblock Asynchronous group mutual exclusion.
\newblock {\em Distrib. Comput.}, 13(4):189--206, 2000.

\bibitem[MN98]{MN98}
M.~Mizuno and Nesterenko.
\newblock A transformation of self-stabilizing serial model programs for
  asynchronous parallel computing environments.
\newblock {\em Information Processing Letters}, 66(6):285--290, 1998.

\bibitem[NA02]{NA02}
M~Nesterenko and A~Arora.
\newblock Stabilization-preserving atomicity refinement.
\newblock {\em Journal of Parallel and Distributed Computing}, 62(5):766--791,
  2002.

\bibitem[NS93]{NS93}
Moni Naor and Larry Stockmeyer.
\newblock What can be computed locally?
\newblock In {\em STOC '93: Proceedings of the twenty-fifth annual ACM
  symposium on Theory of computing}, pages 184--193, New York, NY, USA, 1993.
  ACM Press.

\bibitem[Tel88]{T88}
G~Tel.
\newblock Total algorithms.
\newblock In Vogt FH~(ed) Springer, editor, {\em Concurrency 88}, volume LNCS
  335, pages 277--291. Springer-Verlag, 1988.

\bibitem[Tel04]{Tel94}
G.~Tel.
\newblock {\em Introduction to Distributed Algorithms (Second Edition)}.
\newblock Cambridge University Press, 2004.

\end{thebibliography}
\end{small}

\newpage

\appendix

\section{Self-stabilization of the layer clock}
We apply the convergence stair method~\cite{D00}.%  \cite{GM91}. 
\begin{lemma}
The predicates $WU_{1},WU_{2}$ and $WU$ are closed.
\end{lemma}

\begin{proposition}
\label{pro:dc_1 }
The clock \emph{C}$_{1}$ stabilizes to $WU_{1}$.
\end{proposition}

\begin{proof}
Let $e=\gamma _{1}.....\gamma _{k}....$ be a maximal execution. Assume that $e$
is finite. Then, the last state $\gamma _{l}$ is a deadlock.  So, the clock $\emph{C}%
_{1}$ is stabilized, otherwise there should exist a process for which $CA_{1}
$ or $RA_{1}$ is enable. We suppose now that $e$ is not finite.\ The
projection $e_{1}$ of $e$ on the registers $r_{1}$ is an execution of the
clock \emph{C}$_{1}$. If $e_{1}$ is finite, then in the last state $\emph{C}%
_{1}$ is stabilized for the same reasons than above. If $e_{1}$ is not
finite, then $e_{1}$ is an infinite execution of $\emph{C}_{1}$, so from~\cite{BPV04b} there is a state which is in $WU_{1}$.
\end{proof}

\begin{proposition}
\label{pro:dc_2 }
The clock \emph{C}$_{2}$ stabilizes to $WU_{2}$ .
\end{proposition}

\begin{proof}
Let $e=\gamma _{1}.....\gamma _{k}....$ be a maximal execution. We can
assume from Proposition \ref{pro:dc_1 } that $\gamma _{1}\in WU_{1}$.\ So while $C_{2}$
is not stabilized, $C_{1}$ does not execute any action. So  \ the projection 
$e_{2}$ of $e$ on the registers $r_{2}$ is an execution of the clock \emph{C}
$_{2}$. If $e_{2}$ is finite, then in the last state $\emph{C}_{2}$ is
stabilized otherwise there should exist a process for which $CA_{2}$ or $%
RA_{2}$ is enable. . If $e_{2}$ is not finite, then $e_{2}$ is an infinite
execution of $\emph{C}_{2}$, so from \cite{BPV04b} there is a state which is in $WU_{2}$.

\end{proof}

From Proposition~\ref{pro:dc_1 } and Proposition \ref{pro:dc_2 } we deduce the corollary:

\begin{corollary}
\label{cor:WUstab-appendix}
The layer clock stabilizes to $WU$.
\end{corollary}

\begin{proposition}[No starvation]
Once stabilized, the clock \emph{C}$_{1}$ increments infinitely often.
\end{proposition}

\begin{proof}
Let $e=\gamma _{1}.....\gamma _{k}....$ be a maximal execution. We can
suppose from Corollary~\ref{cor:WUstab-appendix} that $\gamma _{1}\in WU$.\ Assume that for a
process $p$, action $NA$ is  executed only a finite number of time. Then the clock of
each process is executed only a finite time, so $e$ is finite. But in the
last state of $e$, minimal processes according to the precedence
relationship are enable, which is a contradiction.

While $C_{2}$ is not stabilized, $C_{1}$ does not execute any action. So \
the projection $e_{2}$ of $e$ on the registers $r_{2}$ is an execution of
the clock \emph{C}$_{2}$. If $e_{2}$ is finite, then in the last state $%
\emph{C}_{2}$ is stabilized otherwise there should exist a process for which 
$CA_{2}$ or $RA_{2}$ is enable. If $e_{2}$ is not finite, then $e_{2}$ is
an infinite execution of $\emph{C}_{2}$.  From~\cite{BPV04b} there is a state
which is in $WU_{2}$.
\end{proof}

\section{Correctness Proofs of Subsection~\ref{sub:SSLRA}}

\paragraph{$\varrho$-LRA.}

 In order to proof liveness and no lockout property,  we lifts the main clock, using the lifting construction defined in section~\ref{sec:Unison_analysis}.  
We use the same notation, except that the register $r$ becomes the register $r_1$. As in Section~\ref{ssec:infimum_calcul}, $\delta=\rho+1$.

\begin{lemma}[No lockout]
In each phase $\left[ C_{U\delta  },C_{U\delta  +\varrho }\right] $, there is
at least one process which is elected and which increments.
\end{lemma}

\begin{proof}
The set of processes is finite. So there is an infimum $\left( p.r_{2}, p.v\right) $ among the processes, 
according to  the total order relationship $\preceq $.  $p$ is elected during the phase $\left[ C_{U\delta  },C_{U\delta  +\varrho }\right] $ and $p.r_2$  increments.
\end{proof}

\begin{lemma}[Liveness]
\label{lem:liveness}
Every process has the privilege infinitely often.
\end{lemma}

\begin{proof}
It is sufficient to prove that each process $p$ has the privilege at least
one time. Assume that a process $p$ has never the privilege. 
Let $Pot_{p}=\sum\limits_{q\in V}\delta_{( p,q)}$ be the some of the delays from $p$ to
each $q$ in $V$. This quantity is an integer, and from the assumption that $p$ has never the privilege,  $Pot_{p}$ is  strictly increasing.
So $Pot_{p}$ is not upper bounded. But this quantity is upper bounded by
 $\left| V\right| D$. This is a contradiction. We deduce the lemma.
\end{proof}

\paragraph{$\varrho$-LME.}

From Section~\ref{subsec:usedc}, both the no lockout and liveness properties are verified. It remains to show the safety property.

\begin{lemma}[Safety]
If the process $p$ has the privilege, then no process at distance less than
or equal to $\varrho $ from $p$ has the privilege simultaneously. 
\end{lemma}

\begin{proof}
Assume that $p$ has the privilege in the phase $\left[ C_{U\delta },C_{U\delta  +\varrho }\right]$, 
 it enters critical section  when its register  satisfies $\widetilde{p}.r_{1}=U\delta  +\varrho $. 
Any other process $q$ at distance less than or equal to $\varrho$ from $p$ has not the privilege in the phase 
$\left[ C_{U\delta  },C_{U\delta  +\varrho }\right]$.  So if $q\in B_{\varrho }\left( p\right) $ has the
privilege simultaneously, it does not have the privilege in the same phase. The absolute value of the delay,
according to the first clock, between the two processes is less than or equal to $\varrho$.  But when $p$ enters critical section and while $p$ is in critical section:
 $\widetilde{q}.r_{1}\in \left\{ U\delta  +\varrho -\varrho ,...,U\delta  +\varrho +\varrho
\right\} =\left\{ U\delta  ,...,U\varrho +2\varrho \right\} $, 
and $U\delta  +2\varrho  <\left( U+1\right) \delta  +\varrho $.
 We deduce that $q$ can enter in critical section simultaneously if and only if $\widetilde{q}.r_{1}=U\delta  +\varrho $, thus in the same phase as $p$, which leads to a
contradiction. 
\end{proof}

%From these properties we conclude:
%
%\begin{theorem}
%The $\varrho -LME$ algorithm solves the \emph{Self-Stabilizing} $\varrho$-LME problem.
%\end{theorem}

\paragraph{$\varrho$-Group mutual exclusion.} 

Again, from Section~\ref{subsec:usedc}, both the no lockout and liveness properties are verified. 
The next lemma directly follows from the construction of $cond$:
\begin{lemma}[Safety]
If the process $p$ and the process $q$ have the privilege simultaneously and are at distance less than or equal to $\varrho $,
the requested resources are in the same group.
\end{lemma}
%From liveness, no lockout property and from this lemma we conclude:
%
%\begin{theorem}
%The $\varrho -$group mutual exclusion algorithm solves the \emph{Self-Stabilizing }$\varrho -$\emph{group mutual exclusion }problem.
%\end{theorem}

%\newpage
%\appendix
%\tableofcontents

\end{document}